**Absolute ethanol intake drives ethanol preference in *Drosophila***

Scarlet J. Park[a,b], William W. Ja[a,b*]

[a]Skaggs Graduate School, The Scripps Research Institute, Jupiter, FL 33458
[b]Department of Neuroscience, The Scripps Research Institute, Jupiter, FL 33458
* To whom correspondence should be addressed. Email: wja@scripps.edu.

**Abstract**

Factors that mediate ethanol preference in *Drosophila melanogaster* are not well understood. A major confound has been the use of diverse methods to estimate ethanol consumption. We measured fly consumptive ethanol preference on base diets varying in nutrients, taste, and ethanol concentration. Both sexes showed ethanol preference that was abolished on high nutrient concentration diets. Additionally, manipulating total food intake without altering the nutritive value of the base diet or the ethanol concentration was sufficient to evoke or eliminate ethanol preference. Absolute ethanol intake and food volume consumed were stronger predictors of ethanol preference than caloric intake or the dietary caloric content. Our findings suggest that the effect of the base diet on ethanol preference is largely mediated by total consumption associated with the delivery medium, which ultimately determines the level of ethanol intake. We speculate that a physiologically relevant threshold for ethanol intake is essential for preferential ethanol consumption.

**Introduction**

*Drosophila melanogaster* is a powerful model for human behaviors and disorders, including alcohol use. Numerous reports have detailed factors that alter ethanol consumption, including sex (Park et al., 2018), mating status (Shohat-Ophir et al., 2012), previous exposure to ethanol (Devineni and Heberlein, 2009, Ja et al., 2007, Peru Y. Colón de Portugal et al., 2014), and genotype (Devineni et al., 2011, Sekhon et al., 2016). Despite progress in characterizing fly responses to ethanol, factors that mediate consumptive ethanol preference are less well understood; some have speculated that experience (Peru Y. Colón de Portugal et al., 2014) or the caloric value of ethanol (Pohl et al., 2012) drive voluntary ethanol consumption, while others disagree (Devineni and Heberlein, 2009, Xu et al., 2012).

The ongoing debate surrounding ethanol preference stems in part from differences in the assays used to quantify feeding behavior. Previous studies of ethanol preference have employed diverse methods that vary in aspects such as requiring the animals to be starved prior to the experiment or the mode of food delivery (Park et al., 2018, Devineni and Heberlein, 2009, Xu et al., 2012, Devineni et al., 2011, Peru Y. Colón de Portugal et al., 2014, Ja et al., 2007, Pohl et al., 2012, Shohat-Ophir et al., 2012). A commonly used method is the capillary feeder (CAFE) assay, which excels in highly quantitative measurements of food consumption (Ja et al., 2007). However, the CAFE assay requires animals to feed on liquid diets from the tip of a glass capillary, which has led some to hypothesize that CAFE feeding induces undernutrition in animals and greatly confounds previous ethanol preference studies (Park et al., 2018). A recent paper using a qPCR-based assay (BARCODE), in which flies feed on oligonucleotide-labeled solid (agar-based) food to determine relative intake, proposed that consumptive ethanol preference is sexually dimorphic—well-nourished males show no ethanol preference, whereas female preference is unaltered by hunger (Park et al., 2018). However, there it is difficult to directly compare the results from CAFE assays—based on absolute consumption of liquid diets—and those from the BARCODE assay—based on relative consumption of solid food.

To reconcile the discrepancies in observed ethanol preference across reports and identify a common factor that reliably predicts ethanol preference, we modified a radiolabeled feeding assay (Deshpande et al., 2014) to measure solid food consumption of flies offered a choice between different food sources. This method allows quantitative assessment of voluntary, preferential ethanol consumption on solid food that is easily accessible in standard fly vials—avoiding potential artifactual hunger from experimental paradigms (Devineni and Heberlein, 2009, Ja et al., 2007, Park et al., 2018, Peru Y. Colón de Portugal et al., 2014, Pohl et al., 2012, Shohat-Ophir et al., 2012). We report that the dietary medium into which ethanol is mixed is a major determinant of ethanol preference and that the effect is not sex-specific. By manipulating total food consumption using non-nutritive tastants to avoid altering dietary caloric content, we show that ethanol preference can be eliminated or induced even when the nutritive value of the base diet remains constant. Correlative analysis of preference for different concentrations of ethanol derived from various diets shows that absolute ethanol intake and total food consumption are stronger predictors of observing ethanol preference than actual caloric intake or the dietary caloric content.

Our findings suggest that the effect of base diet on ethanol preference is largely mediated by total consumption associated with the delivery medium, which ultimately determines the level of ethanol intake, and we speculate that there exists a physiologically relevant threshold for ethanol intake that is essential for preferential ethanol consumption. While these correlative analyses cannot definitively rule out any caloric contribution—from dietary content or consumption—to ethanol preference as previously proposed, the hypothesis that total ethanol intake determines ethanol preference may reconcile previous accounts of ethanol preference.

**Materials and methods**

*Fly stock and culture*

Canton-S flies were reared on a standard cornmeal-sucrose-medium (5.8% cornmeal, 1.2% sucrose, 3.1% active dry yeast, and 0.7% agar (all w/v), supplemented with 1% (v/v) propionic acid and 1% (v/v) methylparaben mixture (22.2% methylparaben, w/v, in ethanol) in an incubator with controlled temperature (25°C), humidity (60%), and light (12/12-hr light/dark cycle) for the entire duration of the experiments. Adult flies were collected 0-2 days after eclosion and sexed two days later under $CO_2$ anesthesia on standard medium (5 or 10 single-sex flies per vial) and transferred to fresh food vials every other day. All experiments were performed using flies that were 4-6 or 5-7 days old.

*Two-choice solid food intake measurement*

Total consumption of radiolabeled medium was measured as described (Deshpande et al., 2014) with modifications. All test diets were prepared one hour prior to the experiment from heated (~60°C) solutions of 0.5% *Drosophila* agar (Apex), sucrose (Sigma), Bacto tryptone (BD Biosciences), sucralose (Alfa Aesar), and/or papaverine (Adipogen). Indicated diets were then diluted with $ddH_2O$ or ethanol to the desired concentration immediately before dispensing into vials. Each vial contained two ~600μL patches of food, one with ethanol and the other without, on opposite sides (Fig 1A). Half of the vials contained 1-2 μCi/mL [α-$^{32}$P]dCTP (PerkinElmer) in the non-ethanol food and the other half in the ethanol-containing food. Once the medium had solidified, flies were transferred to the experimental vials around ZT 3. After 24 hours, flies were collected in empty vials and killed by freezing at -80°C for scintillation counting. Flies from individual vials were submerged in scintillation fluid (ScintiVerse BD Cocktail, Fisher Scientific) and assayed in a multipurpose scintillation counter (LS 6500, Beckman Coulter). Pre-weighed aliquots of non-solidified media were used to calibrate feeding measurements. Each replicate is a measurement of 5 or 10 flies from a single vial.

*Absorption efficiency measurement*

Absorption efficiency was measured essentially as described (Wu et al., 2019). Male adults were provided with ~0.25 mL of radiolabeled medium (5% sucrose + 0.5% agar (w/v)) containing 0%, 5%, or 10% (v/v) ethanol in a small plastic cap (5.5-mm in diameter) to maximize excreta collection. After 24 hours, flies were removed, the cap containing solid food was discarded, and excreta from the walls of the tube were collected with 1 mL 1×PBST. $^{32}$P radioactivity in fly bodies ($cpm_{fly}$) and excreta ($cpm_{excreta}$) were separately quantified, and absorption efficiency was calculated as $\frac{cpm_{fly}}{cpm_{fly}+cpm_{excreta}}$.

*Statistical analysis*

Consumption amounts of food with or without ethanol were compared using Welch's *t* test or two-way analysis of variance followed by Tukey-adjusted *post hoc* comparisons of least-square means. Preference index (PI) was calculated as $PI = \frac{\bar{X}_E - \bar{X}_C}{\bar{X}_E + \bar{X}_C}$, where $\bar{X}$ is the mean consumption of the food with ($\bar{X}_E$) or without ($\bar{X}_C$) ethanol. Standard deviation of preference index was estimated by propagating the standard deviations of the consumption of either food. The calculated PI was compared to 0 with one-sample Student's *t*-test. To dissociate the likelihood of both null and alternative hypotheses, we used the Bayesian framework for performing correlation analyses between PI and the variables of interest. A Bayes factor (BF) between 1/3 and 1 was considered anecdotal evidence for the null hypothesis ($H_0$: $\rho = 0$), a BF between 3 and 10 a moderate evidence for the alternative hypothesis ($H_1$: $\rho \neq 0$), and a BF between 10 and 30 a strong evidence for $H_1$ per convention (Lee and Wagenmakers, 2014, Jeffreys, 1961). Region of practical equivalence (ROPE) decision criterion was used for equivalence testing: reject $H_0$ if the 95% high density interval (HDI) falls completely outside the ROPE; accept $H_0$ "for practical purposes" if the HDI falls fully inside the ROPE; remain "undecided" otherwise (Kruschke, 2018). When there was more than an anecdotal evidence for the existence of an association (BF < 1/3 or >3), simple linear regression analyses were used to calculate the x-intercept after checking the fitted model for linearity, normality, and homoscedasticity of the residuals. Caloric values of the base diets (sans ethanol) were estimated assuming 3.87 kcal/g for sucrose and 4 kcal/g for tryptone. Statistical analyses were performed using the R statistical software (R Core Team, 2018) and the RStudio environment (RStudio Team, 2016) with lsmeans (Lenth, 2016), lmSupport (Curtin, 2018), psycho (Makowski, 2018), and BayesFactor (Morey and Rouder, 2018) packages, except the manual calculations of *t* scores for the one-sample Student's *t*-tests.

*Data availability*

The complete raw data and their analyses supporting the findings of this study can be downloaded from https://github.com/HungryFly/JaLab/raw/master/publications/ethanol_JEB/SI_dataset.xlsx.

**Results and discussion**

We utilized a radiolabeled feeding assay (Deshpande et al., 2014) to measure consumption when flies were presented with a choice between solid food with or without ethanol in a standard *Drosophila* vial (Fig. 1A). Flies had access to both diets, and the radiotracer was mixed into one food patch in each vial. Ethanol had negligible effect on the near-complete absorption of the radiolabel, confirming that the radiolabel feeding assay can be used to accurately measure the consumption of food with or without ethanol (Fig. 1A). On a tryptone-sucrose (2 and 5 % w/v, respectively) base diet, both males and females consumed greater amounts of food with 5% ethanol than without and showed significant ethanol preferences (Fig. 1B). In partial agreement with a previous report (Park et al., 2018), ethanol preference was dependent on the base diet—a 3-fold concentrated medium eliminated consumptive preference for ethanol-spiked food—in both sexes (Fig. 1C). To further dissect the effect of nutrient concentration on ethanol preference, we next measured ethanol preference in males using a sucrose-concentration series. There was a significant sucrose concentration × ethanol interaction on consumption; flies ate more of the ethanol-containing than the ethanol-free food and showed significant ethanol preference on the lower concentrations of sucrose but not on the highest concentration tested (Fig. 1D).

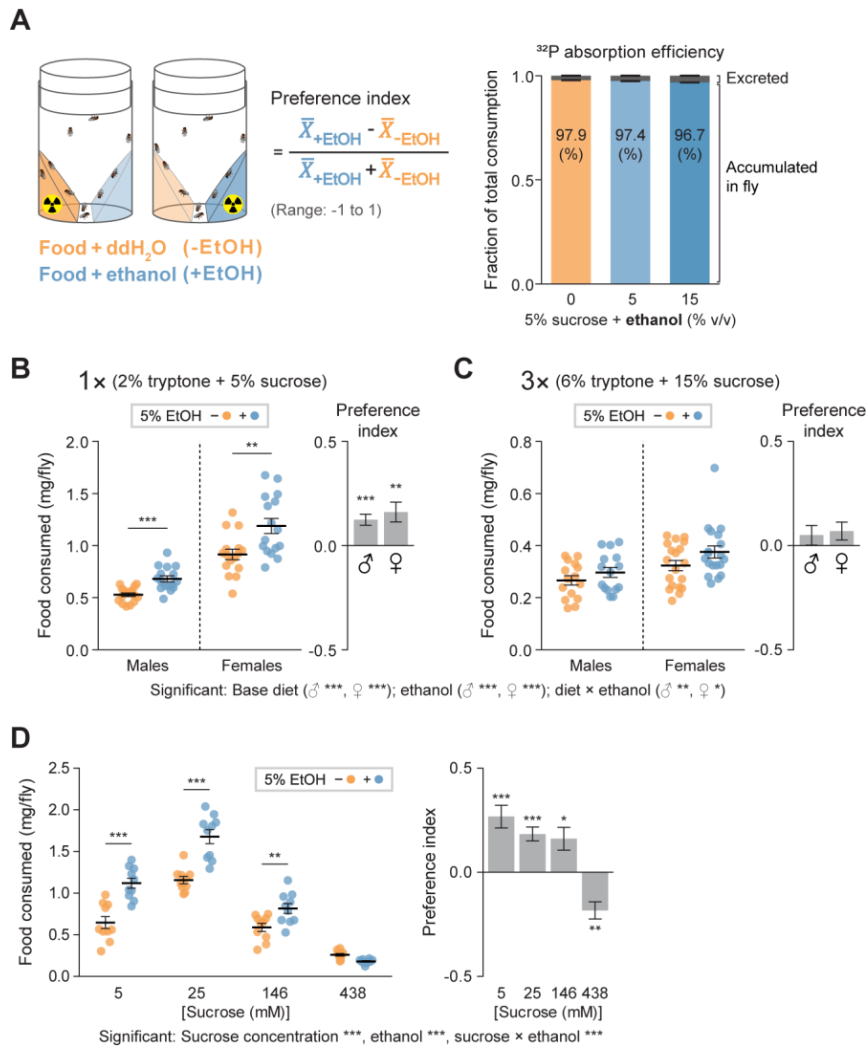

**Figure 1. Both sexes show diet-dependent ethanol preference.**

(A) Schematic of the assay for consumptive ethanol preference. Flies have access to medium with and without ethanol, and only one of the food patches is radiolabeled in each vial. Absorption efficiency of the radiolabel mixed into the diet is ≥ 96.7 % and not substantially compromised by ethanol. (B, C) Both males and females display preference for ethanol on diluted (B), but not concentrated (C), medium. $N$ = 15-19 per condition. (D) Males display ethanol preference only on lower concentrations of sucrose. $N$=10 per condition. PI, preference index. Measurements reflect 24-hour consumption in Canton-S flies. Data shown are mean ± SEM. *, $p<.05$; **, $p<.01$; ***, $p<.001$.

Previous studies have concluded that the display of ethanol preference specific to lower concentration diets indicates that undernutrition underlies ethanol preference (Park et al., 2018). However, since flies can adjust their feeding behavior to compensate for diet concentration, dilutions of a dietary medium do not necessarily produce a similar degree of change in caloric intake—and the consequential hunger state—of the animals (Fig. 1B-D, Table 1) (Deshpande et al., 2014, Carvalho et al., 2005, Ja et al., 2009). Moreover, absolute ethanol intake is inextricably tied to diet; if the concentration of ethanol mixed into the diet remains constant, increasing nutrient concentration may decrease absolute ethanol intake without altering caloric intake from food. Indeed, dietary dilution had a limited effect on caloric intake yet reversed ethanol preference (Fig. 1B-D, Table 1), suggesting that preferential ethanol consumption may be independent of caloric needs but closely related to total food consumption and thus gross ethanol intake.

**Table 1. Summary of feeding parameters on various media**

| Diet | Total consumption (mg/d/fly) | Intake (µg/d/fly) | | | Diet caloric content (cal/mg) | Caloric intake (cal) | Preference Index | p (PI≠0) |
|---|---|---|---|---|---|---|---|---|
| | | Protein | Carb | EtOH | | | | |
| 2 T + 5 S | 1.21 | 22.70 | 56.74 | 33.25 | 0.26 | 0.31 | 0.15 | *** |
| 6 T + 15 S | 0.56 | 30.27 | 75.68 | 13.86 | 0.74 | 0.41 | 0.06 | |
| 5 S | 1.41 | - | 65.89 | 39.85 | 0.18 | 0.25 | 0.16 | * |
| 15 S | 0.44 | - | 59.13 | 8.39 | 0.52 | 0.23 | -0.18 | ** |
| 0.17 S (5 mM) | 1.77 | - | 2.85 | 55.45 | 0.01 | 0.01 | 0.27 | *** |
| 0.86 S (25 mM) | 2.84 | - | 23.17 | 83.09 | 0.03 | 0.09 | 0.18 | *** |
| 2 T + 5 S (*15% EtOH*) | 0.93 | 17.44 | 43.61 | 75.61 | 0.26 | 0.24 | 0.13 | |
| 15 S (*15% EtOH*) | 0.35 | - | 47.08 | 21.12 | 0.52 | 0.18 | -0.12 | * |
| Agar | 0.28 | - | - | 5.35 | 0 | 0 | -0.20 | |
| 0.5% sucralose | 1.76 | - | - | 48.11 | 0 | 0 | 0.10 | * |
| 2.5% sucralose | 2.21 | - | - | 58.85 | 0 | 0 | 0.08 | * |
| 1.71 S (50 mM) | 2.19 | - | 35.58 | 58.79 | 0.06 | 0.14 | 0.08 | * |
| 1.71 S + 1 mM papaverine | 1.36 | - | 22.15 | 34.09 | 0.06 | 0.09 | 0.01 | |
| 1.71 S + 5 mM papaverine | 0.35 | - | 5.69 | 8.63 | 0.06 | 0.02 | -0.01 | |

All data are from male Canton-S under choice conditions with 5% ddH$_2$O or 5% EtOH, except the diets written in blue (15% EtOH; correspond to the blue data points in Fig. 2C). All diets are in 0.5% agar base. The diets written in red indicate non-nutritive media (correspond to the red data points in Fig. 2C). Abbreviations: EtOH, ethanol; PI, preference index; S, sucrose; T, tryptone; number in front of S or T indicates % (w/v). Preference index ranges from -1 to 1, where 1 = full preference for ethanol-containing food. *, $p<.05$; ** $p<.01$; ***, $p<.001$.

To isolate the effect of total consumption—and absolute ethanol intake—on ethanol preference, we manipulated food intake without affecting caloric value of the diet using non-nutritive tastants. On agar alone, male flies preferred the unadulterated agar over that containing 5% ethanol, contradicting the idea that undernourishment drives ethanol preference (Fig. 2A). Adding sucralose to the agar, thereby sweetening the medium without adding any nutritive value, increased overall consumption and induced ethanol preference (Fig. 2A). Conversely, flies show ethanol preference on a 50 mM sucrose base diet, but adding increasing amounts of a bitter compound, papaverine, to the base diet decreased total consumption and eliminated ethanol preference (Fig. 2B).

To evaluate the relative contributions of the different dietary factors on ethanol preference, we compared the preference index for different concentrations of ethanol observed on various base diets with the dietary properties (caloric content) and consumptive parameters associated with each diet (caloric intake, total food consumption, and ethanol intake) (Table 1). Total consumption and gross ethanol intake were the stronger predictors of preference index; results from Bayesian correlational analyses indicated that greater amounts of overall food consumption and final ethanol intake were moderately or strongly, respectively, associated with higher ethanol preference index (Fig. 2C). Dietary nutrient concentration and the actual caloric intake—which is a better reflection of the internal nourishment state of the animals than the former—were both anecdotally unassociated with ethanol preference index (Fig. 2C).

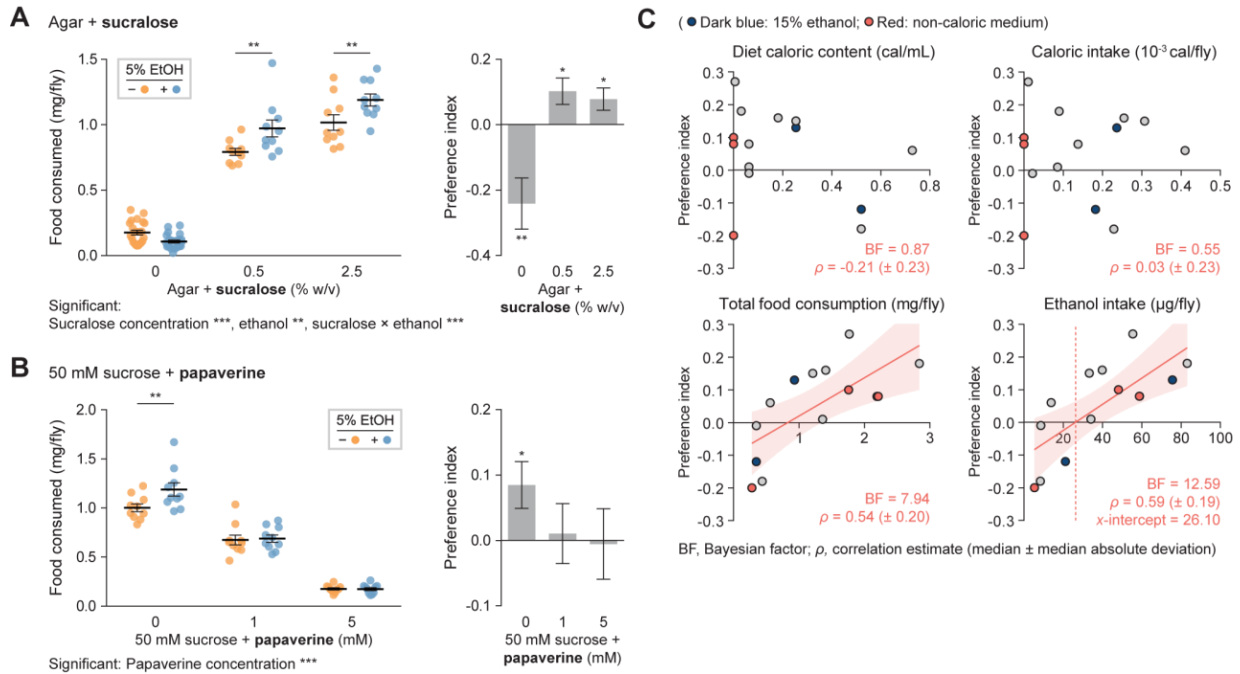

**Figure 2. Absolute ethanol intake predicts display of ethanol preference.**

(A, B) Manipulating total consumption with the non-nutritive tastants, sucralose (sweet, A) or papaverine (bitter, B), alters ethanol preference without affecting the caloric value of the base diet. $N = 22$ (0 % sucralose) or 10 per diet. Data shown are mean ± SEM. *, $p<.05$; **, $p<.01$; ***, $p<.001$. (C) Preference index (PI) for ethanol mixed into various diets correlates with total food consumption and ethanol intake, but not with dietary caloric content. Each point represents PI on base diets varying in composition ($N>10$ per diet). Preference indices are for 5% ethanol except those shown in blue (15%). Red circles represent non-caloric base diets. Results from Bayesian correlation analyses are summarized in each graph. Associations between PI and each of the parameters can be summarized as follows: dietary caloric content, anecdotal evidence for the absence of a negative association; caloric intake, anecdotal evidence for the absence of a positive association; total food consumption, moderate evidence for the existence of a positive association; ethanol intake, strong evidence for the existence of a positive association. Solid line and surrounding shaded area show linear regression and 95% confidence interval.

Although our correlational analysis cannot definitively rule out hunger as a contributor of ethanol preference, the hinted absence of association between dietary caloric content or intake and ethanol preference substantiate the idea that consumptive ethanol preference in flies is independent of caloric contributions. Previous works have shown that flies cannot derive substantial energy from ethanol yet display consumptive preference (Peru Y. Colón de Portugal et al., 2014, Xu et al., 2012), and preference for ethanol consumption—despite being unable to utilize it for energy—has also been reported in honeybees (Mustard et al., 2019).

Human studies of ethanol drinking initiation distinguish between the experiences of the first drink and the first intoxication—which often do not coincide—on their effect on drinking habits in adulthood (Samson, 1987). We speculate that consumptive ethanol preference in *Drosophila* similarly depends on suprathreshold substance intake, as shown in studies of self-administered oral preference for drugs of abuse in rodents (Grim et al., 2018). This explanation is consistent with previous reports that ethanol preference increases with duration and intensity of exposure (Devineni and Heberlein, 2009, Peru Y. Colón de Portugal et al., 2014) and that ethanol preference is attenuated or abolished when the CAFE vials are supplemented with additional food source (Park et al., 2018)—which would effectively decrease overall consumption of the test diets and thus the mixed-in ethanol, making it less likely for animals to ingest enough ethanol to meet the threshold for establishing preference. It was previously shown that internal ethanol concentration of non-starved flies that voluntarily consume ethanol is lower than those required for producing behavioral signs of intoxication such as hyperactivity and loss of postural control (Moore et al., 1998, Wolf et al., 2002). It would be interesting to see how the threshold for internal ethanol concentration that produces voluntary consumption compares to those determined for behavioral intoxication and identify the genes that selectively shift the threshold for consumptive preference, given the obvious relevance for clinical alcohol use disorder.

Contrary to a previous study (Park et al., 2018), we observed similar diet-dependency of ethanol preference in both sexes. However, the sexes likely diverge in their total ethanol intake or internal concentration threshold for developing consumptive preference. Female flies metabolize ethanol more slowly than males (Devineni and Heberlein, 2012), making it plausible that less ethanol intake is required for establishing ethanol preference in females. Taste also likely plays a role in voluntary ethanol consumption. Flies find ethanol unpalatable (Devineni and Heberlein, 2009) and do not readily consume pure ethanol, just as humans use various mixers to mask the flavor of pure ethanol. Studies using other animals have similarly reported that rhesus monkeys will readily self-administer ethanol intravenously but not orally and laboratory mice and rats that are not water- or food-deprived require specific induction protocols to start orally self-administrating various concentrations of ethanol (Meisch, 1977).

Nonetheless, the most parsimonious explanation is that total food consumption and the accompanied net ethanol intake reliably predict development of consumptive ethanol preference. It then follows that variables that affect feeding—including dietary properties (e.g., palatability, nutrition) and experiment duration—or ethanol accumulation (e.g., duration and intensity of ethanol exposure, metabolism) would be expected to have an indirect effect on consumptive ethanol preference. Our work adds to the growing body of evidence that shows how changes in feeding behavior contribute to diverse phenotypes including aging, development, metabolism, and sleep (Keebaugh et al., 2017, Keebaugh et al., 2018, Park et al., 2017, Yamada et al., 2017, Ja et al., 2009). Future studies, especially those focused on the psychopharmacological properties of ethanol, should test for the possibility that changes in total food intake contribute to addiction-like phenotypes.

## Acknowledgments

We thank Dr. Erin S. Keebaugh for providing helpful comments on this manuscript.

## Competing interests

The authors declare no competing interests.

## Funding

This work was funded by the National Institutes of Health (R56AG065986, to W.W.J.). S.J.P. received support from The Celia Lipton Farris and Victor W. Farris Foundation Graduate Student Fellowship.

## Data availability

The complete raw data and their analyses supporting the findings of this study are included in the Supplementary Information.